\documentclass[sn-mathphys]{sn-jnl}

\usepackage{booktabs} 
\usepackage{makecell}
\usepackage{subcaption}
\usepackage{adjustbox}
\usepackage{multirow}
\newcommand{\ffbox}[1]{%
	{
		\setlength{\fboxsep}{+1\fboxrule}
		\fbox{\hspace{1.2pt}\strut#1\hspace{1.5pt}}
	}
}

\jyear{2021}%
\usepackage[numbers]{natbib}

\begin{document}

\title[Op2Vec]{Op2Vec: An Opcode Embedding Technique and Dataset Design for End-to-End Detection of Android Malware}

\author*[1]{Kaleem Nawaz Khan}\email{kaleemnawaz@uetmardan.edu.pk}

\author*[1]{Najeeb Ullah}\email{najeebullah@uetmardan.edu.pk}
\author*[2]{Sikandar Ali}\email{sikandar@uswat.edu.pk}
\author[3,4]{Muhammad Salman Khan}
\author[5]{Mohammad Nauman}
\author*[6]{Anwar Ghani}\email{anwar.ghani@iiu.edu.pk}

\affil[1]{Department of Computer Science, University of Engineering and Technology Mardan, Pakistan}

\affil[2]{Department of Computer Science and Technology, University of Swat, Pakistan}

\affil[3]{AI in Healthcare, Intelligent Information Processing Lab, National Center of Artificial Intelligence, UET Peshawar, Pakistan}

\affil[4]{Department of Electrical Engineering (JC), University of Engineering and Technology Peshawar, Pakistan}

\affil[5]{Department of Computer Science, National University of Computer and Emerging Sciences, Peshawar, Pakistan}

\affil[6]{Department of Computer Science and Software Engineering, International Islamic University Islamabad, Pakistan}

\abstract{Android is one of the leading operating systems for smart phones
	in terms of market share and usage. Unfortunately, it is also an appealing
	target for attackers to compromise its security through malicious applications. To tackle this issue, domain experts and researchers are trying different techniques to stop such attacks. All the attempts of securing Android platform are somewhat successful. However, existing detection techniques have severe shortcomings, including the cumbersome process of feature engineering. Designing representative features require expert domain knowledge. There is a need for minimizing human experts' intervention by circumventing handcrafted feature engineering. Deep learning could be exploited by extracting deep features automatically. Previous work has shown that operational codes (opcodes) of executables provide key information to be used with deep learning models for detection process of malicious applications. The only challenge is to feed opcodes information to deep learning models. Existing techniques use one-hot encoding to tackle the challenge. However, the one-hot encoding scheme has severe limitations. In this paper, we introduce; (1) a novel technique for opcodes embedding, which we name Op2Vec, (2) based on the learned Op2Vec we have developed a dataset for end-to-end detection of android malware. Introducing the end-to-end Android malware detection technique avoids expert-intensive handcrafted features extraction, and ensures automation. Some of the recent deep learning-based techniques showed significantly improved results when tested with the proposed approach and achieved an average detection accuracy of 97.47\%, precision of 0.976 and F1 score of 0.979.}

\keywords{Andriod malware detection, opcode embedding, end-to-end learning, deep learning}

\maketitle

\section{Introduction}\label{Intro}

\par Mobile technology has shown exponential growth in recent past. Mobile devices are the best source to manage our day-to-day communications. These mobiles accompany us in all our movements. Use of these devices allow us to handle most of our very important activities, social networking, payments and banking, with ease. Due to high growth rate, the mobile platforms are highly targeted, and severely infected with malicious applications~\cite{R60}. Attackers look for possibility of exploiting mobile platforms through various techniques. There are cyber criminals as well as hackers, sponsored by states, doing research in order to find schemes for possible attacks against mobile platforms for their better interest. According to International Data Corporation (IDC)~\cite{R15}, mobiles have surpassed PCs in terms of preferred devices that can be used to access internet and other possible services. IDC also states that number of mobile users will cross 91 million mark over the coming four years.
\par Malware is a short form used for malicious software. It is a program or software on any system that is not intentionally installed by end user or system administrator. There are various types of malware for different tasks and purposes~\cite{R66}. Behavior of a malware can range from being a very simple piece of annoyance, such as pop-up advertisements to severe actions which may be much more damaging and harmful~\cite{R64}, such as stealing important systems' passwords or secret data and other more severe actions. They may be used for infecting other machines having very confidential and secret information, over the network~\cite{R65}.
\par Among mobile platforms, Android is one of the most prevalent platforms for smart phones nowadays. It has seen exponential growth with market share of 82.6\%, and having several millions of mobile applications in various markets~\cite{R15}. It is a very rich platform in terms of availability of various functionalities to its users. Unfortunately, it has been observed that smart phones with Android operating system are targeted more often than any other platform by the security attackers~\cite{R64}, and it is very severely infected by malicious softwares. Unlike other mobile platforms, Android allows easily to install applications from sources without clear verification, such as third-party markets, whose sole purpose is to bundle and distribute mobile applications with malwares, assisting attackers in different kind of tasks~\cite{R67}. According to a report~\cite{R28}, the number of Android malicious applications will cross 3.8 million mark at the end of this year. Keeping this evidence in mind, there is a need for techniques and solutions to limit the production of malwares on different Android markets. Large body of research is involved to overcome the situation~\cite{R56},~\cite{R57}. Researchers are trying to find out smart ways for automated detection of malicious applications.
\par Android applications can be analyzed in two ways: either performing static analysis~\cite{R69} or dynamic analysis~\cite{R70}. In static analysis, application is studied in its static position. Its behaviors, i.e. code patterns, requested permissions, relationships with other applications, intent filters and other features, are analyzed. On the other hand, in dynamic analysis, the application is studied, and analyzed during its running state. Dynamic aspects, such as observation of system calls, dynamic loading of code segment, invocations of API calls, are analyzed. Dynamic analysis is performed mostly in a control environment that is named as sandbox~\cite{R73}. All the relevant operations of state of execution are monitored, such as sending SMS messages, storage reading and connection to remote servers.
\par The conventional Android malware detection pipeline is to take Android applications and use domain expertise to extract handcrafted features from a set of applications. Dynamic and static features are extracted for dynamic and static analyses respectively. The features are then used to train machine learning algorithms to produce trained models to classify and detect Android malware. Common classifiers used for Android malware detection are, support vector machine (SVM), decision tree (DT), k-nearest neighbors (KNN), random forest (RF), neural networks (NN) and k-means clustering. Recently, deep neural networks are getting attention for Android malware detection. Studies, such as Droid-Sec~\cite{R88}, DeepDetector~\cite{RM121},~\cite{RM118}, and ~\cite{RM119} are using deep learning approach to detect Android malware. Unlike handcrafted features extraction for conventional machine learning algorithms, deep learning has a very strong and unique approach to automatically extract deep features and learn classification patterns. 
\par
\begin{figure}[t]
	\centering
	\includegraphics[scale=0.5]{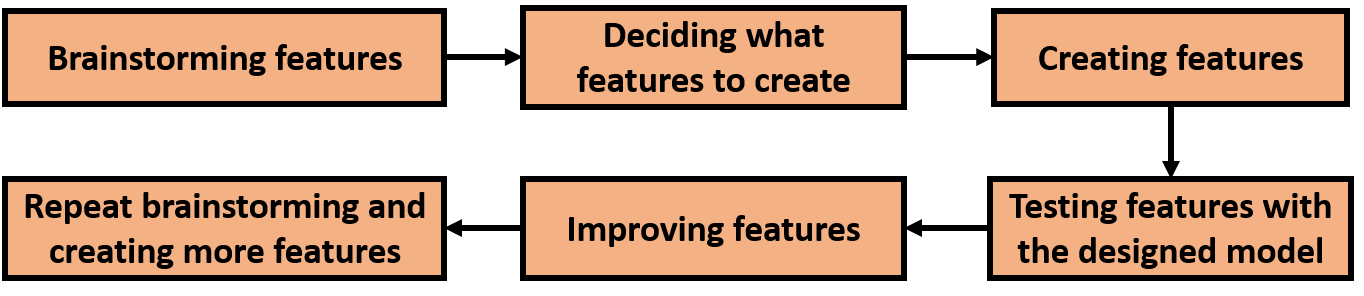}
	\caption{Features engineering process}
	\label{Efeatures}
\end{figure}

 All the conventional machine learning as well as the deep learning techniques studied in the existing literature work well with reasonable accuracy, but the problem is that these techniques rely on engineered handcrafted features. Even the deep learning techniques are trained with handcrafted features. Features engineering is a cumbersome and a very lengthy process, that requires domain knowledge. The feature engineering process is depicted in Figure~\ref{Efeatures}. Domain knowledge and domain experts are required to perform brainstorming of features to decide what features to create. The created features are then tested with the experimentation model. Features are tuned where required and the complete feature engineering cycle is repeated if necessary. In most cases the malware designers are required to design the representative features. The domain experts and the available known malware designers are limited in number. That is why there is a need for making a system that can replace this lengthy and cumbersome features engineering process and incorporate end-to-end learning. The major problem is that we don't have any dataset publically available for deep learning algorithms to learn end-to-end, i.e. extract deep features instead of designing handcrafted features. In end-to-end learning the algorithm learns deep features instead of taking engineered features. So, the gap in the current research is to develop a dataset for end-to-end learning of Android malware and allow deep learning algorithms to be trained on the dataset and detect Android malware with minimum human expert intervention. Another problem in the existing solutions is that they use one-hot encoding to feed information to deep learning models. One-hot encoding creates severe problems and sometimes it becomes infeasible to be used with deep learning techniques. These limitations are discussed in detail in the coming sections. We need to devise an alternative that outperforms one-hot encoding.
\par Our first contribution concerns the development of a novel dataset for end-to-end detection of Android malware. The dataset can be used to learn useful patterns and information from the Android source code. We have not only developed the dataset but have also presented the design process and techniques involved in the dataset development. End-to-end learning minimize human experts' intervention in designing/developing representative features and circumventing the handcrafted features extraction process. The second contribution of this study is learning Op2Vec. Op2Vec learning process employs a machine learning algorithm to learn meaningful vector representations from opcodes of Android source files. All the limitations of existing embedding techniques motivated us to develop a novel encoding technique which we term  ``Op2Vec''. After learning Op2Vec and dataset development we also validate the dataset by feeding it to a deep neural network. Based on the above discussions the main contributions are summarized as follows,
\begin{enumerate}
	\item Op2Vec for learning opcode-vector embedding
	\item Development of a dataset for end-to-end learning based on learned Op2Vec embedding and feeding the developed dataset to deep learning models for learning patterns and insights (circumventing feature engineering).
\end{enumerate}
\par In this section, after a detailed introduction to the problem area and different techniques, the relevant background of Android malware detection and motivation for the proposed approach is discussed in Section II. Section II focuses on malware detection techniques along with details of machine learning models, more importantly, brief introduction and working paradigm of deep learning. Moreover, in last part of the section, word embeddings technique, its specifications, and its relevance to our novel Op2Vec technique are discussed. In Section III, methods of our research and the sub tasks of the proposed approach are listed along with brief discussion, more importantly the word embedding model, Skip-gram, is presented in depth with its working fashion. Section IV discusses the experiments performed and results of the experiments along with detailed analysis. This section also covers, how to use the designed dataset with deep neural networks for end-to-end learning. Section V provides comparison of our proposed dataset with other state-of-the-art datasets and discusses how the use of Op2Vec makes it better than existing datasets. Section VI concludes the paper, and provides further extensions in this part of research.

\section{Background and Motivation}
\subsection{Android Malware Analysis}
\par Due to the severity of Android malware, there are a hundreds of approaches for its detection and classification. One approach is to perform static analysis of the Android application. In static analysis the executable code of the Android applications is examined to determine the data flow, control and representative pattern without running the executables. Another approach is to examine an Android application through dynamic analysis. In dynamic analysis the execution of an application is monitored by inspecting execution behaviors. Both static and dynamic analysis studies are discussed in the next two subsections.
\subsection{Dynamic Analysis Techniques for Android Malware}
\par Due to the complexity of dynamic analysis there are very few techniques conducting dynamic analysis for Android malware detection. A very popular technique is DroidScope~\cite{R10}. It is an Android malicious applications' analysis engine, based on emulation, which performs dynamic analysis of Android applications. Its specialty is to reconstruct both OS and Java level semantics seamlessly and simultaneously. Riskranker~\cite{R12} is an another approach that ensures an accurate and very scalable detection of zero-day android malware. It dynamically analyzes whether a particular application exhibits dangerous behavior. Another similar technique DroidRanger~\cite{R9} performs behavioral foot printing scheme, based on requested permissions, for detection of new samples of unseen Android malware families, and later apply a heuristics based filtering approach for the identification of certain inherent behaviors of unseen malware application families.
\par Android malware detection based on system calls~\cite{R14}, where automatic classification is performed based on tracking system calls. DroidScribe~\cite{R39} is a very recent approach that focuses on dynamic analysis, i.e. runtime behavior. This approach shows how machine learning algorithms can be used to automatically classify Android malware into different malware families by just observing their runtime behavior. It observes, on Android systems, system calls solely does not provide sufficient semantic content to make classification thats why it also uses a light weight VM introspection to reconstruct inter-process communication on Android system for effective analysis. All the discussed techniques in this section have used engineered handcrafted features which are specifically designed for dynamic analysis.
\subsection{Static Analysis Techniques for Android Malware}
\par Static analysis is relatively simple and scalable. There are several techniques that use static analysis for Android malware detection. A set of static features is designed using domain expertise. The features are then utilized for detection of Android malware. An approach, DroidAPIMiner~\cite{R11} statically mines API-level features for the robust detection of malware on Android systems. It aims to provide a lightweight, and robust classifier in order to evade Android malware installation. Another similar technique Droidminer~\cite{R4} uses a behavioral graph to embed abstract malware program logic into a sequence of threat patterns and then apply machine learning techniques to identify and label elements of graph that match already extracted threat modalities. A semantic-based technique~\cite{R1}, that is responsible for classification of Android malware through dependency graphs. It extracts a dependency graph, i.e. weighted contextual API graph, as program semantics for construction of feature sets. Another study reveals that malicious application treats sensitive data differently from benign applications. This feature is used to design a useful technique~\cite{R20} for identification of malware applications. Benign Android applications are mined for their data flow from sensitive sources, and then these flows are compared against those, found in malicious applications, to detect similarity.

\par DREBIN~\cite{R3} is a lightweight method for Android malware detection. It enables identification of malware applications on smart phones directly. It extracts as many features of an Android application as possible, and later the extracted features are embedded into a joint vector space in such a fashion that typical patterns, that are indicative for malicious applications, can be automatically detected and can be used to explain decision making. DREBIN has shown an effective rate of detection of malicious applications on its own dataset. They have designed a dataset and Support Vector Machine (SVM) classifier is used for training on the same dataset.
\begin{table}[b]\hspace*{0.07cm}
	\caption{Features Set for Android Malware Analysis}
	\label{Features}
	\centering
	\resizebox{\columnwidth}{!}{%
		\newcommand{\minitab}[2][l]{\begin{tabular}{#1}#2\end{tabular}} 
		\begin{tabular}{|c|c|} 
			\hline 
			\multirow{4}{1in}{Features for Dynamic Android Malware Analysis} & {Behavioral foot printing}\\ 
			& Dalvik instruction traces\\ 
			& System Calls\\ 
			& API level activity and information leakage\\ 
			& heuristic based filtering\\ 
			& Sensitive data handling\\ 
			& Run time behavior\\ 
			\hline 
			\multirow{4}{1in}{Features for Static Android Malware Analysis} & {Behavioral Graph}\\ 
			& API level features\\ 
			& Weighted contextual API dependency graph\\ 
			& Control flow transition\\ 
			& Permissions\\ 
			& Opcode patterns\\ 
			& Opcode frequency histogram\\ 
			& Manifest and creator information\\ 
			& Information flow\\ 
			& Context, time and connection based network features\\ 
			& Statistical features\\
			& N-gram sequential features\\ 
			& Weighted sum of permissions\\
			& Internet component call graph\\ 
			& Signature based features\\
			\hline 
		\end{tabular}
	}
	
\end{table}
\par A model presented in~\cite{R38} is a n-grams based technique. Sequential features, based on n-grams, are extracted from files' content. It determines patterns of sequential ngrams, then calculation of statistics for the extracted pattern and in the last, classifier is trained to classify malware in different malware families. They have studied three classifiers for this task, i.e. SVM, C4.5 and multilayer perceptron. A method proposed in~\cite{R31} uses three metrics, weighted sum of permissions' subset, a set that consists of combination of permissions and a specific subset of system calls' occurrence. An approach based on permissions' combination is studied in~\cite{R36}. This scheme uses permission information present in the manifest of an Android application. These permissions' combinations are frequently requested by malicious applications and rarely requested by benign applications. Based on the permissions combinations, rules are generated in order to make classification of applications as benign or malicious. This model is then used for classification of unknown applications.
\par An Android applications classification technique~\cite{R85} and other~\cite{R101} use bytecode for detection purpose. The bytecode of an application contains accurate behavior of an application, which can provide enough information about application's intentions. Similarly, malicious applications tend to have same pattern of bytecodes which makes it unique and differentiate it from benign applications.
\par Android file's opcodes are considered as source of information for malware detection. A technique that uses opcodes as features for identification of malware applications is studied in~\cite{R18}. This method is based on state-of-the-art classifiers applied to the frequencies of opcode ngrams. Similarly, a very popular study~\cite{R37} presents a detection mechanism based on features, such as sequences of opcodes combined with machine learning algorithms. As an initial input features, it collects all the possible k-grams in a given set of applications. To determine key relevant features, a selection algorithm is applied and a classifier is trained based upon selected features. Another opcode related study~\cite{R78} applies static analysis over opcodes distribution. In this particular experiment, Android executables are disassembled statically, and the opcodes frequency distribution is extracted. This distribution patterns are then compared with non-malicious executables' distributions. It is noticed that there is a significant difference between these distributions. This feature can be effectively applied to differentiate malware and benign Android applications. Exploitation of opcodes for malware detection is currently a hot topic for Android malware detection problem.

\subsection{Representative Handcrafted Features for Malware Detection}
\par After the extensive study of both dynamic as well as static analysis of Android malware it is concluded that dynamic and static analysis use separate set of features for malware analysis. Features and their importance can be studied and justified using domain knowledge and expertise from computer security domain. Each approach consider different types of features and try to justify their importance and relevance to the problem, there is no standard solution or set of features. The different categories of the features are classified in Table~\ref{Features}.
\subsection{Machine Learning Algorithms for Malware Detection}
\par Once features engineering and extraction step is performed, the next step is usually to apply machine learning algorithms, such as random forest, decision tree, SVM, Bayesian classifiers, KNN, k-means clustering, for malware detection and classification. Each algorithm deals with the extracted features differently and try to learn useful patterns. The parameters of algorithms are tuned considering the nature of features. Studies, such as~\cite{R97}, and~\cite{R98} demonstrate the use of machine learning algorithms for automatic Android malware detection.

\begin{equation}\label{eq:1}
y_{k} =\sum_i {(w_{ki}. x_{i})} \qquad 
\end{equation}
\begin{equation}\label{eq:2}
E_{n} =\frac{1}{2}\sum_k {(y_{nk} - t_{nk})^2} \qquad 
where; y_{nk}=y_{k}(x_{n},w) 
\end{equation}
\begin{equation}\label{eq:3}
\frac{\partial E_{n}}{\partial w_{ji}}
= \left( y_{nj} - t_{nj} \right)x_{ni} 
\end{equation}

\subsection{Deep Neural Networks and Deep Learning}
\par Deep learning or end-to-end learning is a paradigm of machine learning
but very practical~\cite{R47}. It achieves great flexibility and power by learning
to express the world in the form of nested hierarchy of concepts, such
that each complex concept is defined in terms of very simpler concepts.
Concepts that are more abstract are defined in relation to less abstract
concepts. This concept of end-to-end learning is not new but was there for
decades. But recently with very strong hype, end-to-end learning is getting
extraordinary attention. The more prominent contribution of the end-to-end
learning is automatic features engineering~\cite{R48}. This property makes it unique and more powerful than all other existing machine learning techniques.

\par Convolutional Neural Networks (CNNs) is an example of deep neural network, it tries to learn low-level concepts or features such as edges and may be lines in starting layers, then parts and pieces of objects, and then high-level representation of the objects~\cite{R74}. The general problem solving approach of deep learning is different than conventional machine learning algorithms. These algorithms first divide the problem into sub problems, and then after solving all the sub problems the results are complied in a collective form. This is not the case in deep learning because it is end-to-end learning, i.e. just input the raw data and the classification results are collected as output. CNNs have multiple hidden layers where convolution operation is performed, few dense layers near the end of the network, an output and input layer . The input data is fed to the input layer along with random initial weights. The output is received from output layer. All the calculations are performed in hidden layers. Equation~\ref{eq:2} is error function known as mean squared error, where y is the predicted output and t is the target output or may be named as label. We can also consider y as linear combinations of the given inputs as presented in Equation~\ref{eq:1}. Where y\textsubscript{k} is the output, w\textsubscript{ki} is weight and x\textsubscript{i} is the input. After prediction, error is calculated using Equation~\ref{eq:2}, y\textsubscript{nk} is the predicted output while t\textsubscript{nk} is the ground truth. This error function is for some input n, and all possible outputs or labels k. The term x is a particular input having weight w. The calculated error is then back-propagated to hidden layers and weights are updated accordingly. This process is continued till the weights are optimized enough, and the value of error function is converged. Equation~\ref{eq:3} is the error function gradient considering some weight w\textsubscript{ji}. This gradient is used in back-propagation process.
\begin{figure}[t]
	\begin{center}
		\includegraphics[scale=0.5]{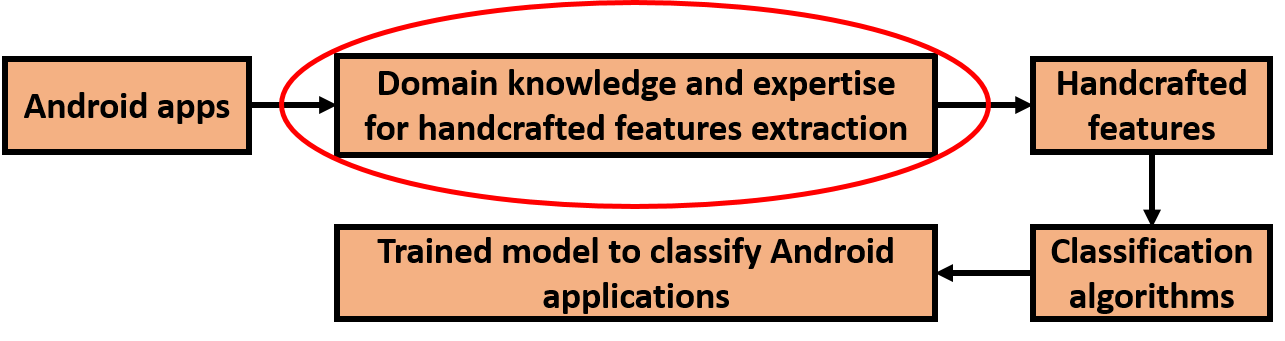}
	\end{center}

	\caption{Conventional model for Android app classification}
	\label{Conventional}
\end{figure}
\begin{figure}[t]
	\begin{center}
		\includegraphics[scale=0.5]{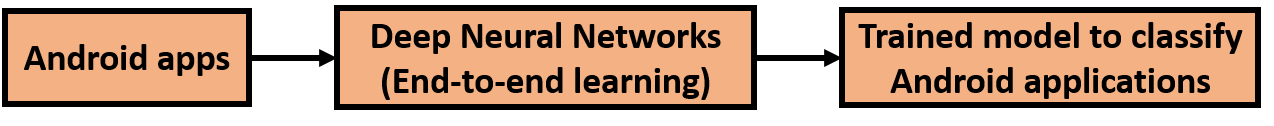}
	\end{center}

	\caption{Proposed classification pipeline}
	\label{Proposed}
\end{figure}
\subsection{Deep Learning Based Analysis for Android Malware}\label{DLBA}
\par There are a number of deep learning based methods for Android malware detection. One of the methods is Droid-Sec~\cite{R88}. It uses 200 different dynamic and static features with deep neural networks for malware detection. DeepDetector~\cite{RM121} is another approach, where eight different set of features are considered to be used with deep neural networks. A multimodal deep learning method is proposed in~\cite{RM118}. The technique uses opcode features and method API features with deep neural networks. Another approach~\cite{RM119}, which extracts sequence of API method calls and manually categorize the dangerous APIs. The categorized API sequences are used to train deep neural networks. In~\cite{RM122} authors use five different set of features to perform classification using deep learning. Droiddetector~\cite{RM125} is considering three set of features, i.e. required permission, sensitive API and dynamic behavior, with deep neural networks.

\subsection{Motivation for the Proposed Approach}
\par Deep learning methods are popular because of deep features extraction. All the existing deep learning based techniques discussed in section~\ref{DLBA} have used handcrafted features for automated Android malware detection as depicted in Figure~\ref{Conventional}. These techniques do not exploit the deep feature extraction nature of deep learning methods rather use engineered handcrafted features. To hand-design, an effective feature is a lengthy process. This approach is very costly, lengthy and require intensive domain knowledge and expertise as already discussed in section~\ref{Intro}. There is a need for making this features extraction process automatic and reduce human experts' intervention. Aiming at new applications, deep learning enables to acquire new effective feature representation from the available dataset for training. The major difference between deep learning and conventional methods is that deep learning automatically learns features from big data, instead of adopting handcrafted features as shown in Figure~\ref{Proposed}, which mainly depends on prior knowledge of designers and is highly impossible to take the advantage of big data.
\par Recent studies, such as~\cite{R18},~\cite{R37}, and~\cite{R78} offer solid justification that opcodes distribution and sequences provide important information to differentiate a malware from trusted Android application. The problem is how to feed the opcodes information to deep neural networks as deep neural networks only work with numerical data. Another problem for Android applications classification and malware detection is that there are only a limited number of datasets. The latest publically available dataset is Drebin~\cite{R3}. Drebin and all other existing datasets have android apks structured directories and have such a format that they cannot be used to perform automatic malware detection using deep learning models~\cite{RM126}.
\par First we need to develop an opcode embedding technique to feed the opcodes to deep neural networks. One solution is one-hot encoding~\cite{RM115}~\cite{RM111}~\cite{RM116}. But this solution has shortcomings, i.e. one-hot encoding leads to inefficiency for high dimensional input data and it doesn't capture the semantic relationship. There are extensive studies~\cite{R52M}, and ~\cite{RM123} that strengthen the concept that vector embedding technique outperforms the conventional one-hot encoding solution. This motivates us to develop a vector embedding technique for opcodes which we name "Op2Vec". And secondly, based on the learned vector embeddings, develop a dataset that will be used for end-to-end detection of Android malware.
\begin{figure*}[t]
	\centering
	\includegraphics[scale=0.45]{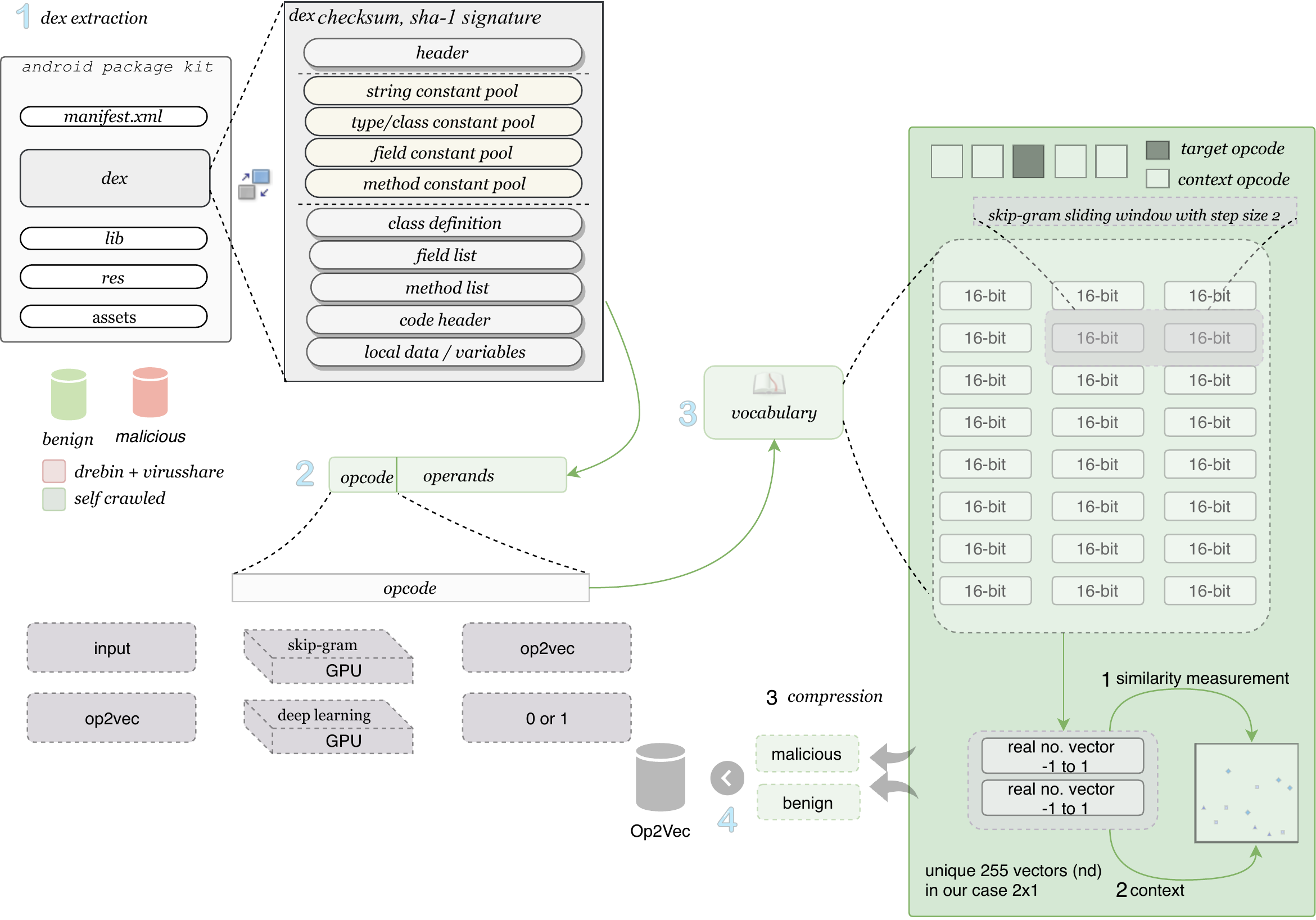}
	\caption{Step-wise workflow of Op2Vec model and dataset design}
	\label{WF}
\end{figure*}
\subsection{Word Embeddings}
Word embedding is a technique used for embedding words into vectors. It does this task in such a fashion that the syntactic and semantic relationship between words are always preserved. Word2Vec is one of the very important sub-tasks for most of the applications of natural language processing (NLP). Word embedding is the collective name for a set of feature learning and language modeling techniques in the field of NLP where words and phrases from the available vocabulary are mapped to vectors of real numbers. Our work is inspired from TWEET2VEC~\cite{RM111}, ATTACK2VEC~\cite{RM112} and ASM2VEC~\cite{RM113} where the authors have applied a word embedding technique to encode tweets, attacks and assembly language functions into vector. Word embedding is considered as a very recent version of these embeddings which are relatively dense and have low-dimensionality~\cite{R105}, which is very efficient in terms of computation. Get motivated from all these studies we are applying the Word2Vec to opcodes which we name Op2Vec, to learn vectors for opcodes. There are two common word embedding models, i.e. Continuous bag-of-words (CBOW) and Skip-gram model. Because of the two severe limitations of CBOW~\cite{RM133}, i.e. ignoring the order of words and ignoring the semantics of words, we will be considering the Skip-gram model for our solution.

\section{Proposed Methodology}\label{mthd}
In this section we present details of the Op2Vec learning process and the designed
dataset. The dataset consists of opcode sequences of Android applications. Initially the dataset contains 28,570 Android applications. More applications can be easily added in future to confirm the robustness of the proposed technique. The dataset design process consists of five phases. Pictorial view of all the steps involved in the proposed design process is shown in Figure~\ref{WF}. After the collection of Android apks from different online Android Play Stores, the first phase is, extraction of Dalvik Executable (.dex) files from apks. The second phase is, extraction of instructions from executable files. All the instructions are processed and only opcode sequences are extracted. In the third phase, opcode sequences of different files are combined to learn Op2Vec, i.e. opcode embeddings. In the very last stage, for each single file, opcodes are replaced by its corresponding vectors, learned in phase four to finalize the dataset. The final step is to feed the dataset to deep neural networks to ensure its validation for end-to-end learning.
\subsection{Collection of Benign Applications}
Benign applications are those applications, that are solely designed for harmless and smooth fulfillment of user requirements. The main objective of these applications is to
meet the user requirements. Some Android benign applications are free
and others are paid, available in online markets. Because of DRM
restrictions, we have mostly collected free Android applications for our dataset.
Around 16,240 free benign Android applications are downloaded from
Amazon Appstore, SlideME, 1Mobile Market and Google Play. A script\footnotemark \footnotetext[1]{https://github.com/KaleemFAST/playstore-scraper-php.git} is designed to download applications from all the available sources.
\subsection{Collection of Malicious Applications}
Android applications, that are
solely designed to meet the developer interests at the cost of harming application's users. Malicious payload is that part of application that carries malicious behavior. There are two main objectives~\cite{R75}, that mostly encourage the design of malicious applications: (1) without the user intention, triggering the malicious payload execution again and again for maximum benefit. (2) escape from
detection in order to have maximum life till the fulfillment of interests. We have collected around 12,330 malicious applications of different malware families, i.e. Androidbox, AnserverBot and 12 other families.
\subsection{Dex Files Extraction}
The Android APK consists of files; classes.dex, resources.arsc, AndroidManifest.xml and subfolders; lib, assets, res and META-INF. The file structure of Android apk is depicted in Figure~\ref{WF} step 1. All the files and folders have necessary information regarding apk file. We are interested in Dalvik executable file, i.e. classes.dex. We extract this file from apk using a tool, named apktool\footnotemark \footnotetext[2]{https://ibotpeaches. github. io}. Apktool is a reverse
engineering tool for Android apks. A script\footnotemark \footnotetext[3]{https://github.com/KaleemFAST/Android\_End2End\_dataset\_design.git} is designed, and each apk file is unzipped through apktool. All the extracted classes.dex files are
collected.  
\begin{figure}[t]
	\centering
	\includegraphics[scale=0.24]{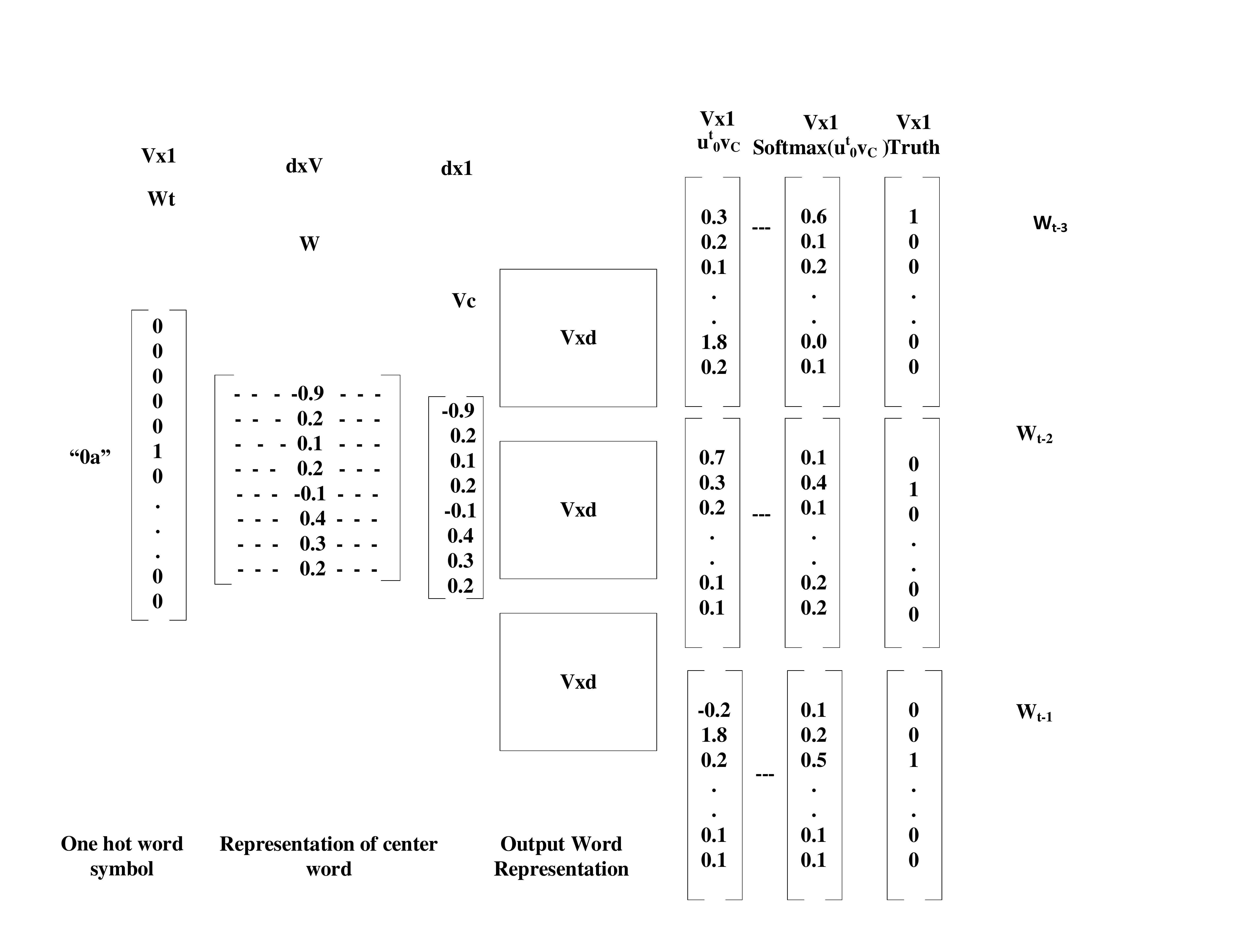}

	\caption{Skip-gram model}
	\label{SGM}
\end{figure}
\subsection{Opcodes Extraction}\label{OE}
Dex file is the executable file for Android
applications. Structure of dex file is depicted in Figure~\ref{WF} step 1. We have
used a tool dedexer\footnotemark \footnotetext[4]{https://sourceforge.net/projects/dedexer/}, i.e. a disassembler tool for Android dex file. We get a dex.log file that produces all the necessary information of different portions of dex file. Using information given in dex.log file, the desired
code section of dex file is filtered. From the code section of dex file, all the
opcode sequences are extracted to another file. This process is repeated for
all the collected dex files. Now we have files containing opcode sequences
of dex files.

\subsection{Need for Opcode Embeddings}\label{NE}
\par Opcodes representation is in textual form, i.e. \emph{05}, \emph{e2}, \emph{03}, \emph{87} etc. There are two severe and main problems with this type of representation, 1)
we can see order relationship among these opcodes, i.e. \emph{05} is greater
than \emph{03}. In reality, there is no such relationship, i.e. opcodes are not comparable with each other and thus are not ordered~\cite{R106}. If we feed this
data to the deep neural network in this form, the network may consider
this relation as a feature (because of feature engineering nature of deep
neural networks), this reduces accuracy and the learning process may be
misled~\cite{RM114}. A technique is needed that should change this representation in such a fashion that assists the learning process and preserves the opcodes identity, but still break the unnecessary ordering, and 2) Deep neural networks doesn't work with textual/categorical data directly rather
we have to change or encode its representation to numeric values.
\par The conventional encoding technique used is one-hot encoding~\cite{RM111},~\cite{RM115}, and~\cite{RM116}. This encoding technique has the following limitations;
\begin{enumerate}
	\item In some scenarios one-hot encoding may be useful where the number of categorical variables are limited~\cite{RM115}, ~\cite{RM116}. But when the variables are not limited, it becomes very expensive and lead to inefficiency in most cases~\cite{RM117},~\cite{RM123}. In our case we have 255 different opcodes so each opcode x\textsubscript{i} if embedded into one-hot encoding, it will have size xi $\in $ [0,1]\textsuperscript{255x1}, and for example if a single file has 300000 opcodes it will have 30000 vectors of dimensions 255x1. This data will exponentially grow if the number of files exceed 50,000 figure, i.e. 50,000 x 300,000 vectors of dimensions 255x1. Training deep learning models on high-dimensional data having no spatial structure causes a major computational problem. It implies a network with an input layer of a very huge size, which greatly increases the number of weights, often making the training infeasible~\cite{RM131}.
	\item It does not capture morphological resemblance between categories and it also ignores the semantic relationship between the input categories~\cite{RM117}. Which can be very useful for deep learning models to learn deep features from opcodes arrangement in Android source file~\cite{RM132}.
\end{enumerate} 
\begin{equation}\label{eq:4}
p(w_{c,j}=w_{O,c} \mid w_I)=y_{c,j}=\frac{exp({u_{c,j}})}{\sum_{j'=1}^{V} exp({u_j'})} \qquad 
\end{equation}
We definitely need to have an opcode embedding technique that overcomes all the above issues.
\subsection{Opcode Embeddings using Skip-gram Model}\label{skp}
From the previous section it is clear that we need an opcode embedding technique. That is why we have introduced Op2Vec to get rid of all the listed issues. We have applied the skip-gram word embeddings technique for opcodes encoding. Skip-gram model~\cite{R52M} is a very prominent model in NLP that is used for Word2Vec. The words are embedded into vectors with the intuition that model needs to learn very similar and almost identical vectors for words having similar contexts. The complete architecture of Skip-gram is shown in Figure~\ref{SGM}. Window size is selected based on the problem nature. The input to the network is one-hot vector that represents the input word and the output is also number of one-hot vectors considering window size. While evaluating the trained network on a word given as input, the vectors that are obtained as output are probability distributions for nearby words. Where from nearby words, we mean words lying inside the window selected for a particular vocabulary file given for training. The weights W in Figure~\ref{SGM}, which are learned at the input layer are the embedded representations of all the words in vocabulary file.
These probabilities are calculated using Equation~\ref{eq:4}. In Equation~\ref{eq:4}, w\textsubscript{O} and w\textsubscript{I} represent the ouput and input vectors respectively. V represents the length of the vector. y is a training instance and u is any given vector. In plan english this equation states the prediction probability of a particular j\textsuperscript{th} word of the c\textsuperscript{th} panel, equals to c\textsuperscript{th} output word, i.e. actual value of the output vector index, conditioned on w\textsubscript{I}. This equation decides the index value for a particular word in the output vector. \par We have used the skip-gram to learn Op2Vec. Skip-gram uses word sequences, so for Op2Vec, words are analogous to opcodes. We applied this concept with the intuition that opcodes appear in the same context must have similar vector representations. Word2Vec is analogy to Op2Vec, i.e. learning Op2Vec, encodes opcodes in such a manner that opcodes having similar semantics are assigned nearly identical vector representations.
\section{Experimental Setup and Results}
\par There are three main experiments that are carried out to justify the efficacy of the proposed approach. One is learning Op2Vec. The second one is the dataset development based on the learned Op2Vec, i.e. opcode embeddings for end-to-end learning of Android malware. And the third experiment is to feed the designed dataset to deep neural networks to validate the claim that the dataset can be used for deep learning based analysis of Android malware. In Figure~\ref{WF}, the gray boxes and cubes from left to right depict the Op2Vec learning process. The skip-gram model takes input and uses GPU facility to learn Op2Vec. These Op2Vec will be used with deep learning models to perform end-to-end learning for Android malware detection and classify Android apps as benign 0 or malicious 1.  
\subsection{Op2Vec: Learning Opcode Embeddings}
\par Words analogy to opcodes is considered in order to apply word embeddings technique to opcode embeddings. Same steps and process of skip-gram model, used for word embeddings in section~\ref{skp}, are applied for opcodes. After the learning phase, opcodes are encoded into vector representations. This process consists of four subtasks which are listed as follows,
\subsubsection{Preprocessing Phase}\label{preprocessing}
\par In the preprocessing phase we consider 5,000 Android applications' dex files, 3,000 benign and 2,000 Malicious files out of the total 16,240 benign Android applications and 12,330 malwares respectively, for development of vocabulary file to train our model. All the opcodes are collected into a single file. Now this file is considered as vocabulary file for the learning phase. This vocabulary file has to be fed to Op2Vec model for training. At the very start we aren't able to feed the opcodes directly to the neural network. The reason is that opcodes are represented in hexadecimal notations, i.e., strings, thats why we binarize this input with one-hot encoding technique. As there are total 255 opcodes, so the length of one-hot input vector is 255x1. For a particular opcode \emph{0a} the one-hot vector is depicted in Figure~\ref{SGM}. For \emph{0a} the corresponding entry in the vector is fixed as 1 and all the other 254 entries are zeros. The output of this model network is a vector of same size. It should also have 255 components. Every entry of the resultant output vector is the probability of an opcode selected randomly in the vicinity of the input opcode.

\begin{figure}[b]
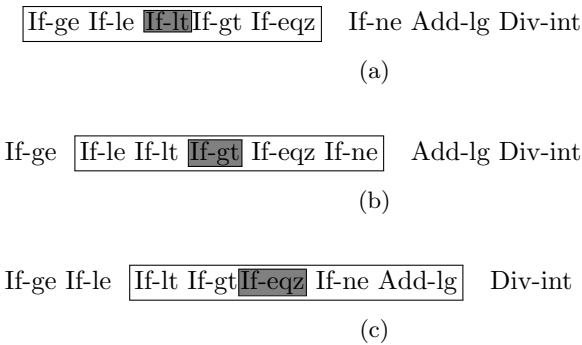

	\begin{subfigure}[t]{0.99\textwidth}
		\hspace{0.9cm}
		{\ffbox{If-ge If-le \fcolorbox{black}{gray}{If-lt}If-gt If-eqz} If-ne Add-lg Div-int}
		\caption{}
		\label{fig:a}
	\end{subfigure}%
	
	\vspace{0.4cm}
	\begin{subfigure}[t]{0.99\textwidth}
		\hspace{0.9cm}
		If-ge\ffbox{If-le If-lt \fcolorbox{black}{gray}{If-gt} If-eqz If-ne} Add-lg Div-int
		\caption{}
		\label{fig:b}
	\end{subfigure}%
	\vspace{0.4cm}
	\begin{subfigure}[t]{0.99\textwidth}
		\hspace{0.9cm}
		If-ge If-le\ffbox{If-lt If-gt\fcolorbox{black}{gray}{If-eqz} If-ne Add-lg} Div-int 
		\caption{}
		\label{fig:c}
	\end{subfigure} 
	\vspace{0.2cm}
	\caption{sliding window over vocabulary opcodes} \label{fg:A}
\end{figure}

\subsubsection{Parameters Setting for Training Phase}
\par Op2Vec embeddings are learned using skip-gram model. It is a neural network based model having all the hidden, input and output layers. All neurons in the hidden layers are without the activation functions, but all the neurons in output layer use the softmax. Softmax is a type of regression used for multi-class classification. In softmax, for a given input X, a designed hypothesis try to estimate \( P(y=k \mid x) \), which is the probability for each value of \(k= 1,...,K\) . This value of k denotes label of a particular class, so essentially the function gives us the probability of a particular input to be in any class k. The network is trained on pairs of opcodes. The input vector is one-hot representation of the input opcode. The output, that is also in the form of one-hot vector, is all the opcodes inside the window except the input opcode, and we call it training output opcodes.
\par For computational simplicity and better visual representation we trained our network for opcode vectors having two dimensions. We already know that the input vector size is 255 and we have selected the dimensions count as 2, so the representation of the hidden layer weight matrix is going to be in the form of matrix with 255 columns and 2 rows. The ultimate goal of this setup is to learn the weight matrix. The output vectors are thrown out once we are done with learning. The network is trained to do the task, i.e. given any specific opcode in the middle of opcodes sequence and randomly pick one opcode from the vicinity, the model tells us the probability for every opcode in the vocabulary to be that opcode we have selected randomly. The vicinity or nearby term is used because skip-gram model use
the window as a parameter in its algorithm, typically the window size parameter is set to 5 as recommended in the original documentation of skip-gram. Window size 5 means 5 opcodes ahead and 5 opcodes behind the central opcode. We have hyper parameter, i.e. window size. For our problem, we fixed this hyper parameter to its default value as 5.
\subsubsection{Training Phase}
\par In the training phase of learning Op2Vec, the vocabulary file designed in the section~\ref{preprocessing} is used as input to the neural network. A slide window of size 5 is adjusted to slide through all the opcodes in the vocabulary file till the end of the file. The neural network tries to optimize its weight matrix after each iteration to adjust the probabilities at the output layer for the opcodes of same context and semantics. 
\begin{table}[t]\hspace*{0.01cm}
	\caption{Learned Op2Vec: The learned vectors}
	\centering
	\begin{tabular}{lll}
		\toprule
		\textbf{Opcode} & {\textbf {X-axis Value}} & {\textbf {Y-axis Value}}\\
		\midrule
		If-ne & {-0.2729177368} & {-0.0875072266}\\
		\midrule
		If-lt & {-0.3726633597} & {-0.017922292}\\
		\midrule
		If-ge & {-0.6149268202} & {-0.0044448727}\\
		\midrule
		If-gt & {-0.6818177649} & {-0.3873034379}\\
		\midrule
		If-le & {-0.3076827262} & {0.1643184456}\\
		\midrule
		If-eqz& {-0.2591792741} & {0.2236180313}\\
		\midrule
		Mul-int& {0.2114985694} &{ 0.3691054416}\\
		\midrule
		Sub-lg & {0.1262099695} & {0.1640332061}\\
		\midrule
		Div-int& {0.1447551711} & {-0.0522292025}\\
		\midrule
		Add-lg & {-0.0497673974} & {-0.2025787514}\\
		\midrule
		Div-lg & {0.1181362618} & {-0.0835916175}\\
		\midrule
		Iput-wide & {0.3703214875} & {-0.0470563225}\\
		\midrule
		Iput-byte & {0.3919335594} & {-0.0879191859}\\
		\midrule
		Iput-char & {0.462135628} & {-0.3112477693}\\
		\midrule
		Invoke-static & {-0.278561079} & {-0.3259538566}\\	
		\midrule
		Invoke-super & {-0.6331899455} & {-0.4750899485}\\
		\midrule
		Invoke-virtual & {-0.5944988213} & {-0.5101055075}\\
		\midrule
		New-array & {0.0389557098} & {0.6073178184}\\
		\midrule
		Filled-new-array & {-0.0966243253} & {0.5125404323}\\
		\midrule
		New-instance & {0.0762253855} & {0.4074241374}\\
		\bottomrule	
	\end{tabular}
	\label{Vectors}
	
\end{table}
\par In Figure~\ref{fg:A}\subref{fig:a}, a sliding window of size 2 is shown, i.e. two opcodes before the central opcode and two opcodes after the central one. For instance 5 opcodes in a row are selected, the opcode colored black is the input opcode for the network. After feeding this setup, the network try to learn patterns in the form of statistical information that the number of occurrence of each pair, i.e. central opcode with any other opcode in the mentioned window. In this particular case where opcode \emph{If-it} is the central opcode the pairs are \emph{(If-lt, If-le)}, \emph{(If-lt, If-ge)}, \emph{(If-lt, If-gt)} and \emph{(If-lt, If-eqz)}. Lets say the pair \emph{(If-lt, If-ge)} occured more frequently in the given vocabulary file so when the learning phase is finished and we input the opcode \emph{If-lt} to the network it shows high probability for the opcode \emph{If-ge} in the output vector. The window is slided further to repeat this process for all opcodes in the vocabulary. This sliding procedure is depicted in Figure~\ref{fg:A}\subref{fig:b} and Figure~\ref{fg:A}\subref{fig:c}.
\begin{figure}[t]
	\centering
\includegraphics[scale=0.6]{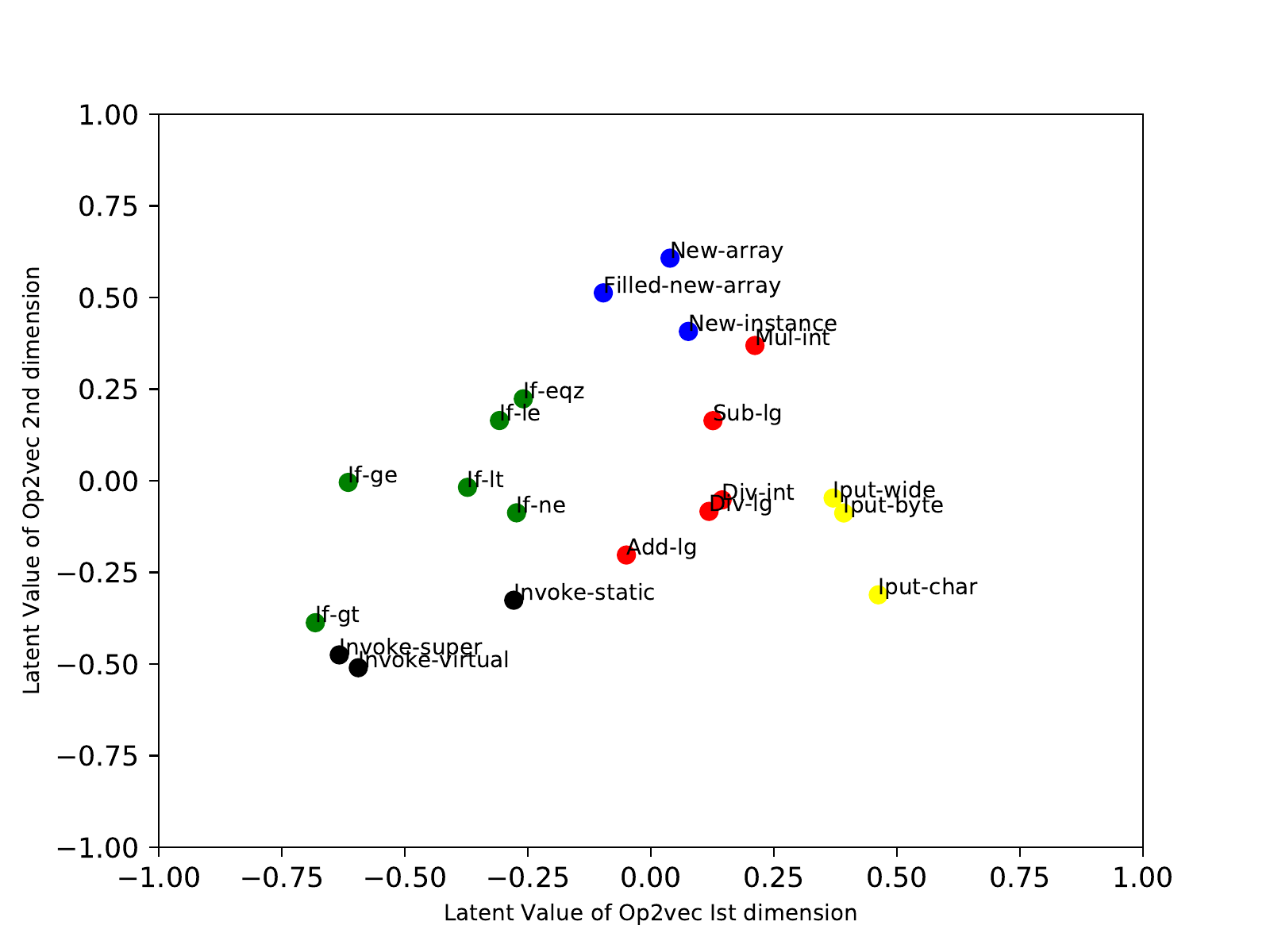}

	\caption{Learned opcode embeddings}
	\label{Groupplot}

\end{figure}
\subsubsection{Learning Op2Vec}
The input vector size is 255x1. The weight matrix for the first hidden layer is of size 2x255 as shown in Figure~\ref{SGM}. The output of the hidden layer is a vector of 2x1. This vector is going to be input for next layer, i.e. output layer, where the weight matrices for this layer is of size 255x2. The output vectors of the network are of size 255x1, i.e. which is the probability distribution of all distinct 255 opcodes. The softmax is applied to the output of each neuron in order to get the values sum up to 1. The columns of learned weight matrix at the first hidden layer is the vector representations of all the 255 opcodes. After the network is trained, when we evaluate the network on a give input opcode, the output vector represents the probability distribution, i.e. a list of values in the form of floating points, not in the form of one-hot vector that we obtain in training phase.

\subsection{Op2Vec Results: Learned Opcode Embeddings}\label{LOE}
\par After completion of the learning process the weight matrix is divided into vectors to get the vector representation of all the 255 opcodes. The values range of all the vectors is in the interval [-1,1]. For some very common opcodes the learned vectors are listed in Table~\ref{Vectors}. These are two dimensional vectors. It can be seen that same categories opcodes are represented by nearly identical vectors. Few of these vectors are plotted in Figure~\ref{Groupplot}. Op2Vec has fixed all the issues of conventional one-hot encoding, discussed in~\ref{NE}, as follows;
\begin{enumerate}
	\item It is very clear from Figure~\ref{Groupplot} that the Op2Vec model learned vectors in such a manner that opcodes, having similar semantics, are represented by almost identical vectors. It is also depicted that opcodes, semantically different, are very apart from each other. If we look at Figure~\ref{Groupplot}, we can clearly see that all the conditional statements~\emph{If-lt},~\emph{If-le},~\emph{If-eqz},~\emph{If-ne},~\emph{If-ge},~\emph{If-gt} are positioned very near in the space, this reveals that semantics are preserved in this sort of learning. Similarly arithmetic opcodes~\emph{Mul-int},~\emph{Sub-lg},~\emph{Div-int},~\emph{Div-lg},~\emph{Add-lg} are separately clustered. Because of the semantic similarity they are almost identical. Same patterns can be observed for the rest of the opcode categories. So the results reveal the fact that Op2Vec has learned embeddings effectively and semantic relationships among opcodes are preserved. Intuitively this is a very useful insight for deep learning models to learn deep features from opcodes arrangement in Android source file. Introducing this relationship among opcodes will enhance the malware detection learning process~\cite{R52M}~\cite{RM123}~\cite{RM124}, which is a contribution in automatic malware detection.
	\item Size of a single one-hot vector is 255x1. Table~\ref{Vectors} shows that Op2Vec embeddings have reduced the 255x1 size of one-hot vectors to 2x1, which significantly decreases computationally complexity~\cite{RM131}.
	\item Op2Vec embeddings have also fixed the limitation of ordering between originally extracted opcodes as there is no such real order in the generated vectors representation.
\end{enumerate}

\subsection{Development of the End-to-end Learning Dataset}
\par Now when the Op2Vec embeddings are successfully learned we can generate the proposed dataset that can be efficiently used for end-to-end learning analysis of Android malware. All the files generated in the section~\ref{OE} are accessed one by one and the opcodes are replaced by their corresponding vector representation learned in section~\ref{LOE}. Thus, the dataset is developed and we claim that this the first ever attempt to develop a dataset that will be used for end-to-end detection of Android malware using deep neural networks. 
\subsection{Feed the Developed Dataset to Deep Learning Models}
The claimed hypothesis that the dataset can be effectively used for end-to-end learning is validated by feeding the dataset to the CNN, i.e. a deep neural network/end-to-end learning model. Each file in the dataset has two dimensional vectors corresponding to each opcode in the original dex file. So each file has two columns and a number of rows. For the network, we consider our input frame consists of two channels as each vector is of two dimensional. Each benign file is assigned label as 0 whereas each malicious file is labeled as 1. Network's setup is all set to process the available files for end-to-end learning. 

\section{Comparison with Existing Datasets}
\par This section draws a comparison among our designed Op2Vec dataset and six other very popular datasets of the malware analysis literature. The comparison is based on the following two parameters.
\subsection {Features (Handcrafted or Deep features)}
\par One of the fundamental limitations of all other available datasets is handcrafted features extraction which are employed for characterizing malware behavior. Our Op2Vec doesn't require handcrafted features. The datasets can be fed directly to the deep neural network for learning deep features.

\begin{table*}[t]\hspace*{-0.03cm}
	\caption{Comparisons with existing datasets}
	\label{Comparisons}
	\centering
	\resizebox{\columnwidth}{!}{%
		\begin{tabular}{lllllll}
			\toprule
			{\textbf{Dataset}}& {\textbf{Files Count}}& {\textbf{Malicious}}& {\textbf{Benign}}& {\textbf{Feature Extraction}}&{\textbf{Features Set}}&{\textbf {Feature Vector}}\\
			\midrule
			{Drebin~\cite{R3}} & {129,013}& {123,453} & {5,560}  & {Handcrafted} & {545,000} & {x\textsubscript{i} $\in$ [0,1]\textsuperscript {545000x1}}\\
			\midrule
			{R. Vinayakumar~\cite{RM127}} & {2296}& {1,609} & {687} & {Handcrafted} & {42} & {x\textsubscript{i} $\in$ [0,1]\textsuperscript {42x1}}\\
			\midrule
			{Wei Wang~\cite{RM128}} & {23,000} & {13,000} & {10,000} & {Handcrafted} & {34,570} & {x\textsubscript{i} $\in$ [0,1]\textsuperscript {413x1}}\\
			\midrule
			{Zhu~\cite{RM129}} & {11,000}& {8,000} & {3,000}  & {Handcrafted} & {323} & {x\textsubscript{i} $\in$ [0,1]\textsuperscript {323x1}}\\
			\midrule
			{VirusShare~\cite{RM108}} & {107,888}& {None} & {107,888} & {Handcrafted} & {482} & {x\textsubscript{i} $\in$ [0,1]\textsuperscript {482x1}}\\
			\midrule
			{Hou~\cite{RM130}} & {5,000} & {2,500} & {2,500} & {Handcrafted} & {1,058} & {x\textsubscript{i} $\in$ [0,1]\textsuperscript {1058x1}}\\
			\midrule
			{AndroZoo~\cite{RM134}} & {3,182,590} & {1,162,150} & {2,020,440} & {Handcrafted} & {50,000} & {x\textsubscript{i} $\in$ [0,1]\textsuperscript {50000x1}}\\
			\midrule
			{Op2Vec Dataset} & {28,570} & {12,330} & {16,240} & {Automated} & {Deep features} & {x\textsubscript{i} $\in$ [-1,1]\textsuperscript {2x1}}\\
			\bottomrule
		\end{tabular}
	}
\end{table*} 

\begin{table*}[t]\hspace*{-0.03cm}
	\caption{Performance enhancement of existing opcode-based techniques}
	\label{Comparisons1}
	\centering
	\resizebox{\columnwidth}{!}{%
		\begin{tabular}{llllll}
			\toprule
			{\textbf{Reference (Year)}}& {\textbf{Features}} & \bfseries\makecell{Deep Learning\\ Technique}& {\textbf{Dataset}}& \bfseries\makecell{Reported Results \\ (Acc)} & \bfseries\makecell{Results with Op2Vec\\ (Acc)}\\
			\midrule
			{Parildi (2021)~\cite{NR1}} & {Opcodes}& \makecell{VirusShare and native \\ Win7 apps} & \makecell{RNN and\\ LSTM} & {\hfill 95\%} & \textbf{\hfill 96.83\%} \\
			\midrule
			{Ren (2020)~\cite{NR2}} & {Opcodes}& \makecell{Google Play store\\ and VirusShare} & {DNNs} & {\hfill 95.8\%} & \textbf{\hfill 97.1\%} \\	
			\midrule
			{Niu (2020)~\cite{NR3}} & {Opcodes}& \makecell{VirusShare, Androzoo \\and Pea Pods} & {LSTM} & {\hfill 97\%} & \textbf{\hfill 98.77\%} \\ 
			\midrule
			{Pekta (2020)~\cite{NR4}} & {Opcodes}& \makecell{Androzoo, Argus group\\ and GooglePlay} & \makecell{RNN and\\ LSTM} & {\hfill 91.42\%} & \textbf{\hfill 96\%} \\ 
			\midrule
			{Zhang (2018)~\cite{NR5}} & {Opcodes}& \makecell{Microsoft in Kaggle 2015\\ and Benign apps} & {ResNet} & {\hfill 98.2\%} & \textbf{\hfill 98.63\%} \\ 	
			\midrule
			{McLaughlin (2017)~\cite{NR6}} & {Opcodes}& \makecell{Genome project,\\ McAfee Labs} & {CNN} & {\hfill 95\%} & \textbf{\hfill 97.53\%} \\
			\bottomrule
		\end{tabular}
	}
\end{table*} 
\subsection {Feeding information to the deep neural network (One-hot encoding or vector embedding)}
\par All the listed datasets and techniques have used one-hot encoding to feed the Android source code information to the classifiers. One-hot encoding has limitations that are discussed in~\ref{NE}. We have proposed Op2Vec which has fixed all the limitations and outperforms one-hot encoding as discussed in~\ref{LOE}.
\par Drebin~\cite{R3} have used a script for automated extraction of different handcrafted features. The features are embedded in one-hot encoding of the form x\textsubscript{i} $\in$ [0,1]\textsuperscript {545000x1}. SVM classifier is trained to classify applications based on their representative feature vectors. Another dataset that is used in~\cite{RM127}. A total of 42 handcrafted features of size x\textsubscript{i} $\in$ [0,1]\textsuperscript {42x1}, are extracted to use with LSTM. Similarly, authors in~\cite{RM128} have used a dataset where 34,570 handcrafted features are extracted. This feature set is reduced to 413 using a feature selection technique. The input feature vector for the classifier is of size x\textsubscript{i} $\in$ [0,1]\textsuperscript {413x1}. Both ~\cite{RM129} and~\cite{RM130} have used 323 and 1,058 features respectively, for machine learning classifiers. Their input feature vectors are x\textsubscript{i} $\in$ [0,1]\textsuperscript {323x1} and x\textsubscript{i} $\in$ [0,1]\textsuperscript {1058x1} respectively. There are a number of datasets available for Android malware analysis and these datasets are only a collection of Android malwares. VirusShare~\cite{RM108} is one of the examples of such datasets. VirusShare has total 107,888 malicious Android applications. Different studies have used this dataset to learn insights for Android malware. A total of 482 features are considered and a feature vector of size x\textsubscript{i} $\in$ [0,1]\textsuperscript {482x1} has used to feed information to machine learning classifiers.
\par All the discussed datasets have a format that cannot be used to perform automatic malware detection using deep learning models. Most of these studies extract features from datasets and these features are used with deep learning algorithms. Deep learning models have a very strong property, i.e. automatic deep feature extraction. They don't need handcrafted features rather raw data is sufficient for training and learning process. Our proposed dataset is designed to exploit deep learning models for deep features. We have designed a dataset that can be fed directly to deep learning models. Unlike other datasets no handcrafted features are required. The input encoded vector size is x\textsubscript{i} $\in$ [-1,1]\textsuperscript {2x1}, which has a very few dimensions in comparison with other existing dataset techniques.    
\par From Table~\ref{Comparisons} it is clear that in terms of features extraction and embedding technique our Op2Vec dataset is far better and adoptable as compared to the other datasets. Performing Op2Vec type embedding technique can reduce dimensions upto 2x1. Adopting this dataset will allow to learn deep features without extraction of hand-crafted features, and with less computational complexity. Rest of the Table~\ref{Comparisons} shows files count in all the three datasets, malicious and benign files count, features extraction method and encoding techniques for features to be fed to machine learning and deep learning algorithms.

\par In order to demonstrate the significance of the proposed approach, some of the recent opcode-based deep learning techniques such as,~\cite{NR1},~\cite{NR2},~\cite{NR3},~\cite{NR4},~\cite{NR5},~\cite{NR6} are trained and tested with the Op2Vec dataset. For fair comparison the same experimental setup is used for the experiments with Op2Vec. It can be seen in Table~\ref{Comparisons1} that performance of the existing techniques significantly improves by incorporating Op2Vec embeddings. All the listed approaches achieve an average accuracy of 97.47\%, where the highest accuracy is achieved with the setup suggested in~\cite{NR5}. It is evident from the results that Op2Vec which incorporates the semantic relationship of opcodes and deep features, enhance the performance of deep learning techniques to detect an Android malware. 
\section{Conclusion and Extensions}
\par Previous work has shown that opcodes of executables have potential information. Opcodes can be considered as features in order to make discrimination between malware and benign Android applications. But these features are very hard to extract or notify. Handcrafted features or information extraction process is very expensive in terms of cost and time. In order to automate the process, effectively identify potential information and extract deep features, end-to-end learning is a perfect solution. This study concerns the learning of Op2Vec and the development of a novel dataset for end-to-end detection of Android malware. Op2Vec learning process employs a machine learning algorithm to learn meaningful vector representations from opcodes of Android source files. The designed opcode embedding technique is used to develop a dataset for end-to-end detection of Android malware. The dataset will be used to learn useful patterns and information from the Android source code. We have not only developed the dataset but have also presented the design process and techniques involved in the dataset development. To the best of our knowledge, we believe this is the first state-of-the-art dataset for end-to-end Android malware detection. The product dataset of this research will be made openly available for further research concerning Android malware detection. Not
only the dataset but also the designed process of the dataset will be made public so that in future, new Android application files can be added to the dataset. This will make our technique robust to deal with newly emerging Android malware.

\par The proposed technique is one of the static Android malware analysis techniques. The limitation of this technique is that it may not capture the dynamic aspects of malware analysis. One of the future directions can be to combine the Dalvik instruction traces technique with the proposed approach to fix this limitation.\footnotetext{https://github.com/KaleemFAST/Android\_End2End\_dataset\_design.git}

\bibliographystyle{unsrt}

\bibliography{sn-article}

\end{document}